\documentclass[12pt,british,english]{article}
\usepackage[T1]{fontenc}
\usepackage[latin9]{inputenc}
\usepackage[a4paper]{geometry}
\usepackage{float}
\usepackage{graphicx}
\usepackage{setspace}
\usepackage[authoryear]{natbib}
\makeatletter

\providecommand{\tabularnewline}{\\}

\newcommand{\lyxaddress}[1]{
\par {\raggedright #1
\vspace{1.4em}
\noindent\par}
}

\@ifundefined{definecolor}{\usepackage{color}}{}

\usepackage{babel}\usepackage[table]{xcolor}

\usepackage{babel}

\makeatother

\usepackage{babel}
\begin{document}

\title{Novel Distances for Dollo Data }

\author{Michael Woodhams$^{1}$, Dorothy A. Steane$^{2}$, Rebecca C. Jones$^{4}$,\\
 Dean Nicolle$^{5}$, Vincent Moulton$^{6}$, Barbara R. Holland$^{1*}$}

\maketitle

\lyxaddress{\emph{1: School of Mathematics and Physics, University of Tasmania,
Private Bag 37, Hobart 7001, Australia}\\
\emph{2: School of Plant Science and CRC for Forestry, University
of Tasmania, Private Bag 12, Hobart 7001, Australia}\\
\emph{3: Faculty of Science, Health, Education and Engineering, University
of the Sunshine Coast Maroochydore DC, 4558}\\
\emph{4: School of Plant Science, University of Tasmania, Private
Bag 55, Hobart 7001, Australia}\\
\emph{5: Currency Creek Arboretum, P.O. Box 808, Melrose Park, South
Australia 5039, Australia}\\
\emph{6: School of Computing Sciences, University of East Anglia,
Norwich NR4 7TJ, United Kingdom}\\
\emph{{*}: Corresponding author. barbara.holland@utas.edu.au; Fax
number (61) 3 6226 2410; Phone number (61) 3 6226 1990}}

\date{Feb 2012 }

\pagebreak{} 
\begin{abstract}
We investigate distances on binary (presence/absence) data in the
context of a Dollo process, where a trait can only arise once on a
phylogenetic tree but may be lost many times. We introduce a novel
distance, the Additive Dollo Distance (ADD), which is consistent for
data generated under a Dollo model, and show that it has some useful
theoretical properties including an intriguing link to the LogDet
distance. Simulations of Dollo data are used to compare a number of
binary distances including ADD, LogDet, Nei Li and some simple, but
to our knowledge previously unstudied, variations on common binary
distances. The simulations suggest that ADD outperforms other distances
on Dollo data. Interestingly, we found that the LogDet distance performs
poorly in the context of a Dollo process, which may have implications
for its use in connection with conditioned genome reconstruction.
We apply the ADD to two Diversity Arrays Technology (DArT) datasets,
one that broadly covers\emph{ Eucalyptus} species and one that focuses
on the\emph{ Eucalyptus} series \emph{Adnataria}. We also \foreignlanguage{british}{reanalyse}
gene family presence/absence data on bacteria from the COG database
and compare the results to previous phylogenies estimated using the
conditioned genome reconstruction approach.

\emph{Adnataria} phylogeny, gene content phylogeny, conditioning genomes. 
\end{abstract}
A fundamental idea in evolutionary biology is that when two species
share a complex trait the most likely explanation of the similarity
is that both species have inherited the trait from a common ancestor.
However, the absence of a particular trait carries far less information,
for instance wings and eyes are complex traits that have been lost
many times independently in different parts of the evolutionary\emph{
}tree of life. As long ago as 1893 Dollo captured this idea in what
is now known as Dollo's Law which states that complex traits may be
gained once somewhere in evolutionary history, and may be subsequently
lost independently many times. 

In this paper we will only consider Dollo models that generate binary
data, recording the presence or absence of some trait. For example
--- does an organism have any genes in a particular gene family? Does
it have some skeletal feature, such as the mammalian inner ear? When
its DNA is digested by a mix of restriction enzymes, is a particular
DNA fragment produced? In reality, determining if a trait is present
or absent can be less clear cut; for example, paralogues can confound
gene presence/absence decisions. In a stochastic Dollo model the gain
and loss of such traits is treated as arising from a simple probability
model. While few situations match Dollo's Law exactly it provides
a useful model in many evolutionary scenarios of interest. Dollo models
have been used to understand gene families \citep{HS,DagnanMartin}
and complex morphological traits \citep{Gould}, and stochastic Dollo
models have been used to study cognates in language evolution \citep{StochasticDollo2,StochDollo3}.

In some of these scenarios such data can have another interesting
property: only traits that are present in particular reference taxa
are visible; or, in other words, the data are censored. This happens,
for example, with array-based studies where a small set of taxa is
used to create a set of traits (i.e. a set of DNA fragments that make
up an array) to which other taxa can be compared. The idea of a reference
taxon also has parallels in the gene-content setting where some authors
have proposed {}``conditioned genome reconstruction'' \citep{LakeRivera},
where one genome is selected as a reference and, for the remaining
genomes, only gene families present in the reference genome are analyzed.

Data thought to follow Dollo's Law traditionally have been analysed
using a parsimony approach \citep{LeQuesne,Farris}. As is normally
the case with parsimony approaches, branch length information is not
taken into account. The use of stochastic Dollo models is relatively
recent \citep{StochasticDollo2,StochDollo3} and, so far, they have
only been implemented in a Bayesian framework. Bayesian methods are
computationally intensive so there is a need for an approach that
is both computationally efficient and statistically consistent. \\
 \textbf{ } This motivated us to develop a distance-based approach
to Dollo data. Our initial motivation was to derive a distance suitable
for phylogenetic analysis of Diversity Array Technology (DArT) data
\citep{DArT_intro} which, by its nature, is censored. On further
consideration we realized that the same formula can be derived directly
from the mathematics of the stochastic Dollo process, or as a limiting
case of the LogDet distance.

In the following sections we begin by deriving the ADD in a general
Dollo context and then show why it also applies to the censored Dollo
model that arises for DArT data. We then describe an intriguing link
to the popular LogDet distances. After introducing a few other binary
distances we present a simulation study that compares the performance
of the new ADD to other binary distances when applied to Dollo data
under a range of censoring schemes. As an illustration of our approach
we apply the new distance to three case studies, two involving DArT
data for \emph{Eucalyptus} species and one using gene content information.
We conclude with a discussion in which we point out some potential
future directions.

\section*{Methods}

\subsection*{Deriving an Additive Distance for the Stochastic Dollo Process}

The description of a stochastic Dollo process which we follow is that
of \citet{HS}, whose discussion is in the context of the gene content
of a genome. This model can be described as a constant-birth, proportional-death
Markov process. New markers are acquired (e.g. genes added to the
genome) at rate $\lambda$, existing markers are independently deleted
with intensity rate $\mu$. $G(t)$ is the set of markers present
at time $t$. We make an initial observation of the set of markers
$G(s)$ at start time $s$. It is assumed that the system is at equilibrium
at time $s$, i.e. it has been evolving by the stochastic Dollo process
for long enough to be independent of initial conditions. The genome
then evolves for a further time $t$ and we observe the marker set
$G(s+t)$. We then put 
\begin{equation}
n_{11}=|G(s)\cap G(s+t)|,\,\, n_{10}=|G(s)-G(s+t)|,\,\, n_{01}=|G(s+t)-G(s)|
\end{equation}
 i.e. $n_{11}$ is the number of shared presences (markers present
in the genome at both time points), $n_{10}$ and $n_{01}$ are the
counts of markers present at one time point but not the other.

Under these circumstances, \citet{HS} prove the following facts: 
\begin{enumerate}
\item $l(s)=|G(s)|=n_{11}+n_{10}$ is Poisson distributed with mean $m=\lambda/\mu$. 
\item If $l(0)$ is chosen according to this equilibrium Poisson distribution,
then the process $G(t)$ is a time-reversible Markov process. 
\item $n_{01}$ is Poisson distributed with mean $m(1-e^{-\mu t})$. 
\item $n_{10}$ is binomially distributed with $l(s)$ trials each with
probability $1-e^{-\mu t}$ of success, where a {}``success'' in
this context is the loss of a marker. 
\end{enumerate}
From 4 we have expected value $E(n_{10}/(n_{11}+n_{10}))=(1-e^{-\mu t})$
and we solve for $t$: 
\begin{equation}
t=-\frac{1}{\mu}\log\, E\left(1-\frac{n_{10}}{n_{11}+n_{10}}\right)
\end{equation}
 and so we get a distance $d$ that is proportional to time given
by 
\begin{equation}
d=-\log\left(1-\frac{n_{10}}{n_{11}+n_{10}}\right)=\log\left(\frac{n_{11}+n_{10}}{n_{11}}\right).
\end{equation}
 Moreover, $d$ is a statistically consistent estimator for $\mu t$
(i.e. $d$ will converge on $\mu t$ as we collect more data.) Note
that by the time reversibility property, we could equally well use
$d=\log((n_{11}+n_{01})/n_{11})$. To make use of all available data
(both $n_{10}$ and $n_{01}$) we average these two distances to give
\begin{equation}
d_{ADD}=\log\left(\frac{(n_{11}+n_{10})(n_{11}+n_{01})}{n_{11}^{2}}\right)\label{eq:add}
\end{equation}
 and call this the Additive Dollo Distance (ADD). \\

\subsection*{Dollo models with censored data}

As mentioned in the introduction, some datasets of interest have an
additional property whereby only markers which are present in the
reference taxon or taxa can be detected. This is referred to either
as {}``censored data'' or as an {}``ascertainment bias''. We want
to extend the ADD distance to cases of both single and multiple reference
taxa. As we will see below, the first case does not require any change
in the formula.

We introduce our approach for censored data that arise in the context
of DArT \citep{DArT_intro}. DArT uses particular restriction enzymes
to create a {}``genomic representation'' (DNA fragments typically
300-1000 bp in length) from one or more reference taxa. The restriction
enzymes used consist of a `rare cutter' (e.g. \emph{Pst} I, that cuts
at sequence CTGCAG) and a `frequent cutter' (e.g. \emph{Bst}N I, that
cuts at sequence TCGA.) Only fragments cut at both ends by the rare
cutter become markers. The fragments are cloned and are then arrayed
onto a glass microscope slide. Genomic representations are prepared
for study samples using the same restriction enzymes. The study samples
are screened (via DNA-DNA hybridisation) with the array. The presence
or absence in the sample of the DArT markers on the slide is recorded
to produce a binary dataset.

A DArT marker can be lost during evolution by mutations which disrupt
the rare cutter target at either end of the marker, or introduce a
new rare or common cutter target within the marker. Once lost, a marker
can only be regained by reversing the mutation which caused loss,
before another loss-causing mutation occurs. This is a rare event,
so we can model DArT marker gain/loss as a Dollo process.

The data from a DArT analysis suffer from ascertainment bias: only
markers which are present in the reference taxon or taxa (from which
the array was prepared) can be detected. This can distort distances,
depending on the proximity of the taxa to the reference. For example,
the Hamming distance 
\begin{equation}
d_{H}=\frac{n_{10}+n_{01}}{n_{00}+n_{01}+n_{10}+n_{11}}
\end{equation}
 will underestimate distances between taxa distant from the reference,
as both taxa will have few markers in common with the reference, hence
$n_{00}$ will be large. This illustrates a general theme: joint absences
($n_{00}$) do not carry the same meaning as joint presences ($n_{11}$)
and should not be accorded equal weight in the distance formula.

The most appropriate way to infer phylogeny from DArT data is not
obvious, and previous authors have taken a variety of approaches,
often analyzing the same data multiple ways. The distance of \citet{NeiLi}
is the most commonly used distance measure for DArT data \citep{JamesEtal,WenzlEtal,XiaEtal}.
It is designed for restriction site data, which DArT does produce,
but for fragments generated by a single restriction enzyme, rather
than the two enzymes used by DArT. Furthermore, the Nei Li distance
does not account for the ascertainment bias caused by observing only
those markers that are present in the reference taxa. Other distances
that have been used are LogDet \citep{JamesEtal} and DICE \citep{MaceEtal}.
Non distance-based methods include maximum parsimony \citep{JamesEtal,SteaneEtal}
and Bayesian analysis \citep{JamesEtal,SteaneEtal}. Sometimes a principal
coordinate analysis is done in place of, or in addition to, a phylogenetic
analysis \citep{JingEtal,JamesEtal,YangEtal}.

Consider DArT data for taxa $A$ and $B$ generated from an array
constructed with markers taken from taxon $R$. The presence/absence
data for $A$ and $B$ are summarized by the counts $n_{00}$, $n_{01}$,
$n_{10}$ and $n_{11}$ (where $n_{01}$ is the number of markers
present at $B$ but absent at $A$, etc.). Consider the unrooted phylogenetic
tree for $A$, $B$, and the reference taxon $R$. We denote the unique
internal node of this unrooted tree $X$. As all markers in the dataset
are present at $R$, and the markers evolved by a Dollo process, any
marker present at $A$ or $B$ must also be present at $X$. In particular,
a marker present at $A$ is present at $X$, and the fraction of these
markers which are lost between $X$ and $B$ is $n_{10}/(n_{10}+n_{11})$.
In view of the arguments above, 
\begin{equation}
E(n_{10}/(n_{10}+n_{11}))=(1-e^{-\mu t_{XB}})
\end{equation}
 where $t_{XB}$ is the time (branch length) between $X$ and $B$.
From this, we find a distance proportional to time 
\begin{equation}
d(X,B)=\log\left(\frac{n_{10}+n_{11}}{n_{11}}\right)\propto t_{XB}.
\end{equation}
 By symmetry, we also have $d(X,A)=\log((n_{01}+n_{11})/n_{11})$
and so 
\begin{equation}
d(A,B)=d(X,A)+d(X,B)=\log\left(\frac{(n_{11}+n_{10})(n_{11}+n_{01})}{n_{11}^{2}}\right)=d_{ADD}
\end{equation}
 and, once again, we arrive at the formula for $d_{ADD}$. (Note that
this formula will also work if one of $A$ or $B$ is the reference
taxon $R$.)

In this derivation, we assume the existence of a point $X$ in the
tree such that any marker present at $A$ or $B$ is present at $X$.
Under a Dollo process, this assumption is valid if there is a single
reference taxon, or if there are multiple reference taxa which are
monophyletic with respect to $A$ and $B$. This will not generally
be the case, and indeed it is not the case for the \emph{Eucalyptus}
dataset \citep{SteaneEtal} we will look at in one of the case studies. 

\begin{figure}[H]
\includegraphics[width=0.6\textwidth]{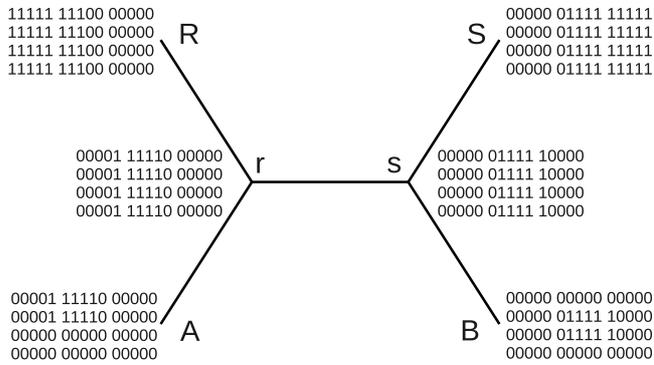} \caption{Multiple reference taxa. The binary strings indicate the presence
or absence of DArT markers. There is a 0.5 probability of marker loss
along any edge. Only markers present at reference taxon R or S are
shown on the diagram. Each character (the 1s and 0s for a given marker)
occurs a number of times proportional to its probability.\label{fig:Multiple-reference-taxa}}
\end{figure}

The potential problem with the ADD for data generated from multiple
reference taxa is illustrated by figure~\ref{fig:Multiple-reference-taxa}.
If we consider only the characters which are present at $R$ (the
leftmost 8 columns at each node), we get the correct answer: $n_{11}=2$,
$n_{01}=2$, $n_{10}=6$, $d_{ADD}=\log_{2}(4\times8/2^{2})=3$ (taking
logarithms as base 2 for convenience.) (There is a 1/8 chance of a
marker at $A$ surviving to $B$, so the $\log_{2}$ path length between
$A$ and $B$ is 3.) However, using the full set of data (which includes
$S$ as an additional reference taxon), now we have $n_{11}=3$, $n_{01}=7$,
$n_{10}=7$ and $d_{ADD}=\log_{2}(10\times10/3^{2})=3.474$. Not all
taxon pairs will suffer this bias, for example in figure~\ref{fig:Multiple-reference-taxa}
if we added a taxon $C$ attaching at point $r$, the distance $d_{ADD}(A,C)$
would be unbiased.

If we know which markers are derived from which reference taxa, then
we can partition the data by marker reference taxon, calculate an
ADD distance matrix for each partition, and then form an average distance
matrix from the partition distance matrices. Under the plausible assumption
that the sampling variances in the partition distances are inversely
proportional to the number of markers in that partition, the minimum
variance estimator is obtained by a weighted average, with weights
proportional to the square root of the number of markers in the partition\emph{.
}We call this weighted average of partitioned distances the Partitioned
Additive Dollo Distance (PADD).

The derivation of the PADD assumes that the markers chosen from each
reference are an independent and unbiased sample from the markers
present in that reference taxon. In practice, this assumption may
fail for two reasons. When constructing the DArT array, we attempt
to eliminate redundancy - so a marker selected for the array from
one reference taxon precludes the same marker being selected from
a second reference taxon in which it may also be present. (Some redundancy
is kept deliberately as an internal control, but omitted from the
final alignments.) Also, markers are chosen which show useful levels
of polymorphism. A marker present in all taxa is not useful, nor is
one which is present only in the reference taxon from which it was
derived.

\subsection*{Links to Conditioned Genome Reconstruction: The Additive Dollo Distance
as a Limiting Case of LogDet}

Another context in which Dollo models may be appropriate is gene-family
presence/absence data. Such data are increasingly available as more
and more genomes are studied. The COG database \citep{cog2} sorts
genes from 50 bacteria, 13 archaea and 3 eukaryotes into nearly 5000
gene families. The gene family presence/absence data have been used
for phylogenetic inference by several authors \citep{LakeRivera,SBS,McCann,Sangaralingam},
but Dollo models have not, to our knowledge, been applied. A problem
noted by previous analysts of these data is that the taxa vary greatly
in the number of gene families they contain. Thus, an unsophisticated
analysis would be biased towards grouping together taxa that have
small genomes. This is somewhat similar to the problem of inferring
a phylogeny in the presence of base-frequency biases, which is an
issue the LogDet distance \citep{logdet,paralinear} was designed
to overcome. When \citet{LakeRivera} sought to apply the LogDet distance
to data from the COG database, they realised they needed to know the
number of shared absences ($n_{00}$). This in turn required the definition
of a {}``universal'' set of gene families. They achieved this by
selecting a `conditioning genome', and taking the set of gene families
present in that genome as being the {}``universal'' set.

In using the LogDet distance, \citet{LakeRivera} explicitly assumed
that gene family presence/absence is a Markov process where a gene
family can disappear from a lineage and later reappear. Frequent horizontal
gene transfer (HGT) was used to justify this assumption. That LogDet
treats shared presences and absences ($n_{00}$ and $n_{11}$) symmetrically,
when physically they have very different meaning, seems to us to be
a potential weakness in their method that has not been commented on
previously.

Lake and Rivera's assumption (that any gene family can be gained by
any taxon at any time) can be thought of as one extreme. The opposite
extreme is to discount the possibility of HGT completely, and adopt
a Dollo model. Consider the standard (non-Dollo) multiple site two
state Markov model: we have $N$ sites evolving independently between
two states (`present' and `absent') by a continuous time Markov process.
The rate for absent$\rightarrow$present transitions is $\alpha$
and for present$\rightarrow$absent transitions is $\mu$. As before,
we sample this process at two different times and get counts $n_{11}$,
$n_{10}$ and $n_{01}$ of shared presences, and of presences at one
time point but not the other. In addition, we get $n_{00}$, the number
of shared absences. In this two state case, the LogDet distance formula
is 
\begin{equation}
d_{LogDet}=-\frac{1}{2}\log\left(\frac{n_{00}n_{11}-n_{01}n_{10}}{\sqrt{(n_{00}+n_{10})(n_{01}+n_{11})(n_{00}+n_{01})(n_{10}+n_{11})}}\right).\label{eq:2 state log det}
\end{equation}
 If we now allow $N$ to vary, we can construct a Markov process so
that in the limit $N\rightarrow\infty$ it becomes a Dollo process.
We require the rate of creation of new markers (sites in the `present'
state) to be $\lambda$, so we set $\lambda=N\alpha$. (The loss rate
$\mu$ is independent of $N$.) As $N\rightarrow\infty$ (and $\alpha\rightarrow0$),
the distributions of $n_{11}$, $n_{10}$ and $n_{01}$ will remain
finite and converge on those for the Stochastic Dollo process described
above. As $n_{11}$, $n_{10}$ and $n_{01}$are finite, we must have
$n_{00}\rightarrow\infty$. Letting $n_{00}\rightarrow\infty$ in
equation \ref{eq:2 state log det} 
\begin{eqnarray}
\lim_{n_{00}\rightarrow\infty}d_{LogDet} & = & \lim_{n_{00}\rightarrow\infty}-\frac{1}{2}\log\left(\frac{n_{00}n_{11}}{\sqrt{n_{00}(n_{01}+n_{11})n_{00}(n_{10}+n_{11})}}\right)\nonumber \\
 & = & \frac{1}{4}\log\left(\frac{(n_{11}+n_{10})(n_{11}+n_{01})}{n_{11}^{2}}\right)\nonumber \\
 & = & \frac{d_{ADD}}{4}\label{eq:limit log det}
\end{eqnarray}
 showing that in this limit, $d_{LogDet}$ is proportional to $d_{ADD}$.

\subsection*{Comparison of binary distances}

Various distances have been defined in the literature for presence/absence
data. We have picked a number of these to compare to ADD including:
\begin{eqnarray*}
\mbox{fractional Hamming distance }d_{H} & = & \frac{n_{10}+n_{01}}{n_{00}+n_{01}+n_{10}+n_{11}}\mbox{ }\\
\mbox{Jaccard distance }d_{J} & = & \frac{n_{10}+n_{01}}{n_{01}+n_{10}+n_{11}}\mbox{ }\\
\mbox{DICE Distance }d_{DICE} & = & \frac{n_{10}+n_{01}}{n_{01}+n_{10}+2n_{11}}\\
\mbox{LogDet distance }d_{LogDet} & = & -\frac{1}{2}\log\left(\frac{n_{00}n_{11}-n_{01}n_{10}}{\sqrt{(n_{00}+n_{10})(n_{01}+n_{11})(n_{00}+n_{01})(n_{10}+n_{11})}}\right)\\
\mbox{Nei-Li distance }d_{NL} & = & -\log(P)\mbox{ where }F=\frac{P^{4}}{3-2P}\mbox{ and }F=\frac{2n_{11}}{2n_{11}+n_{10}+n_{01}}
\end{eqnarray*}
 \citep{BinaryDistanceSurvey,logdet,paralinear,NeiLi}.

\citet{HS} derived a maximum likelihood distance for gene presence/absence
data under a Dollo process 
\begin{equation}
d=-\log\left(\frac{\beta+\sqrt{\beta^{2}+4\alpha_{12}}}{2}\right)\label{eq:Huson Steel}
\end{equation}
 where in our notation $\beta=1-(n_{11}+n_{10}+n_{01})/m$, $\alpha_{12}=n_{11}/m$
and $m=\lambda/\mu$ is the expected number of genes per genome. We
do not know $m$, but if we estimate it by the mean number of genes/markers
at the two taxa $m=[(n_{11}+n_{10})+(n_{11}+n_{01})]/2$ then equation~\ref{eq:Huson Steel}
simplifies to 
\begin{equation}
d=\log\left(\frac{2n_{11}+n_{01}+n_{10}}{2n_{11}}\right).
\end{equation}
 This is a simple transformation of the DICE distance, being $-log(1-d_{DICE})$,
so we name it the Log DICE distance. We can intuitively justify this
transformation, arguing that logarithms correct for multiple events
(e.g. gain, loss, mutation) on the same marker. We can also perform
a similar transformation on the Jaccard distance to create the Log
Jaccard distance $d_{LJ}$. So in summary, in addition to the standard
distances above, we introduce previously unstudied distances: 
\begin{eqnarray*}
d_{LJ} & = & -\log(1-d_{J})=\log\left(\frac{n_{11}+n_{01}+n_{10}}{n_{11}}\right)\\
d_{LDICE} & = & -\log(1-d_{DICE})=\log\left(\frac{2n_{11}+n_{01}+n_{10}}{2n_{11}}\right)\\
d_{ADD} & = & \log\left(\frac{(n_{11}+n_{10})(n_{11}+n_{01})}{n_{11}^{2}}\right)
\end{eqnarray*}
 as well as the composite ADD distance method $d_{PADD}$, defined
above.

\subsection*{The Triangle Inequality and Additivity}

Two important properties of distance functions (i.e. bivariate, non-negative
functions $d(x,y)$ with $d(x,y)=0$ if and only if $x=y$, and $d(x,y)=d(y,x)$
for all $x,y$), are the triangle inequality and additivity. The triangle
inequality states that $d(x,y)\le d(x,z)+d(z,y)$ must hold for all
$x,y,z$ (in which case $d$ is known as a `metric'). Additivity states
that if $z$ was the last common ancestor of $x$ and $y$, and the
sequences evolved independently, then $d(x,y)=d(x,z)+d(y,z)$ should
hold (on average.) The desire for additivity accounts for the presence
of the logarithm function in many phylogenetic distances.

Not all of the distances defined above satisfy the triangle inequality
-- see Table~\ref{tab:Triangle inequality} for counterexamples to
the triangle inequality holding for some of the distances. In Table
\ref{tab:properties of distances} we summarize which distances are
additive and which obey the triangle inequality. The additive Dollo
distance $d_{ADD}$ is additive by construction in the stochastic
Dollo context, and it is a limiting case of the LogDet distance which
is additive \citep{paralinear}. Notably, the only two distances ($d_{H},$
$d_{J}$) known to obey the triangle inequality are not additive,
and the only two distances known to be additive ($d_{logdet}$, $d_{ADD}$)
violate the triangle inequality. Phylogeneticists place greater value
on additivity than on obeying the triangle inequality, as demonstrated
by the popularity of LogDet.

It is worth noting that for $d_{LJ}(x,y)$ and $d_{LDICE}(x,y)$ to
be additive, it is necessary that they go to infinity as the evolutionary
distance between $x$ and $y$ goes to infinity. For the stochastic
Dollo process, $n_{11}=0$ for infinitely separated $x$ and $y$,
so $d_{LJ}$ and $d_{LDICE}$ go to infinity as required, but for
a Markov process where $E(n_{11})>0$ for unrelated $(x,y)$, $d_{LJ}$
and $d_{LDICE}$ will tend to a finite limit. We can generalize the
formulae to correct for this, using 
\[
d_{LJ}(x,y)=-\log(b-d_{J}(x,y))
\]
 where $b$ is the expected value of $d_{J}$ evaluated on uncorrelated/infinitely
separated sequences (and a similar formula applies for $d_{LDICE}$).
For example, for a Markov process where states 0 and 1 are equally
likely at equilibrium, we have $b=2/3$ for $d_{LJ}$ and $b=1/2$
for $d_{LDICE}$.

\subsection*{Simulating censored Dollo data}

The general scheme for our simulations is to create a random tree,
simulate a Stochastic Dollo process along it, select reference taxa
and, finally, select which markers are used to produce an alignment
(on the basis of which markers are present at the reference taxa).

We generate clock-like and non-clock-like trees. For the non-clock-like
trees, we generate the tree topology by a Yule process \citep{yule},
then branch lengths are set so that they are distributed uniformly
between lengths 0.05 and 0.40. For clock-like trees, we generate a
tree by a Yule process with mean branch length 0.1, and repeat this
process until we obtain a tree whose shortest branch is no shorter
than than 0.01. (As short branch lengths are hard to resolve no matter
how good the phylogenetic method, keeping such branches reduces the
contrast between {}``good'' and {}``poor'' methods, which would
make our simulation results harder to interpret.) Our simulated data
are based on both 9- and 15-taxon trees.

For a given simulation run, we specify the expected number of markers
per genome, $m$. As a Dollo process in equilibrium is time reversible
\citep{HS}, we start the process at an arbitrary taxon, with the
number of markers at that taxon drawn from a Poisson distribution
with mean $m$. Then we propagate the set of markers through the tree.
On each branch with length $b$, existing markers are lost with probability
$1-e^{-b}$ each and the number of new markers created has a Poisson
distribution with mean $m(1-e^{-b})$.

We have a number of different models for selecting the markers that
will be included in the alignment to be analyzed, and (for the PADD
method) how the markers are partitioned. 
\begin{description}
\item [{incl1}] (One reference taxon, included.) One taxon is chosen as
a reference. Only markers present at that taxon are selected. There
is only one partition of the markers. 
\item [{excl1}] (One reference taxon, excluded.) As `incl1', except we
discard the reference taxon from the alignment. 
\item [{incl2}] (Two reference taxa, included.) Two taxa are chosen as
references. All markers present at either reference taxon are selected.
For partitioning, a marker which is present at both reference taxa
is assigned randomly to the partition of one of them. Markers which
are present at only one reference taxon go into that taxon's partition. 
\item [{excl2}] (Two reference taxa, excluded.) As `incl2' except we discard
both of the reference taxa from the alignment. 
\item [{all}] (All taxa are references.) All markers are included in the
alignment. Each marker is assigned randomly to the partition of one
of the taxa at which it is present. (We have as many partitions as
taxa.) 
\item [{p2inc}] (Two reference taxa, included, predetermined partitioning.)
Two reference taxa are chosen. Each marker is assigned randomly to
the partition of one of the references. Only markers which are present
at their partition's reference taxon are included in the alignment. 
\item [{p2exc}] (Two reference taxa, excluded, predetermined partitioning.)
As `p2inc', except the two reference taxa are discarded from the alignment. 
\item [{p\_all}] (All taxa are references, predetermined partitioning.)
Each marker is assigned randomly to the partition of one of the taxa.
Only markers which are present at their partition's reference taxon
are included in the alignment. 
\end{description}
For the methods which discard reference taxa (excl1, excl2, p2exc)
we simulate extra taxa at the tree generation stage to account for
the taxa which will be discarded.

In the `incl2', `excl2' and `all' models, if a marker is present in
any reference taxon, it is included in the analysis. This simulates
the circumstance when all possible markers found in the reference
taxa have been included on the DArT array. The `p2inc', `p2exc' and
`p\_all' models simulate the situation where the number of possible
markers is very much greater than the number we can put on the array,
so we get an independent random sampling of markers from each reference.

The models do not all produce the same quantity of data. Compared
to the expected number of markers present at each taxon, the expected
number of markers analyzed is equal for the `incl1', `p2inc' and `p\_all'
models, lower for `excl1' and `p2exc', higher for `incl2', several
times higher for `all', and for `excl2' it depends on the tree, but
for our data is higher on average.

For each scenario (number of taxa, clock like or not, the eight data
selection models) we generate and analyze 5000 random trees.

Once a distance matrix has been calculated, the best tree is found
by minimum evolution, using the program FastME \citep{FastME}. In
addition, we obtain the most Dollo parsimonious tree using the {}``dollop''
program from Phylip \citep{Phylip}. To measure the accuracy of different
methods we record the proportion of splits in the generating tree
that were present in the tree inferred by FastME.

\sloppy Trees in this paper were plotted using the Interactive Tree
Of Life \citep{iTOL2}, and interactive versions may be viewed online
at http://itol.embl.de/shared/mdw. Nexus files containing the raw
data and tree files are available on TreeBase\emph{ }at http://purl.org/phylo/treebase/phylows/study/TB2:S12439.\fussy

A demonstration Perl program, and instructions on its use, for calculating
the ADD, log Jaccard and log DICE distances is included in the supplementary
material.

\section*{Results}

\subsection*{Simulated data}

Table \ref{tab:bad splits} shows the proportion of splits (i.e. edges)
in the reconstructed trees which were incompatible with the true tree.
(Additional tables for differing numbers of characters and number
of taxa are provided in the supplementary material, tables 1--10.)
The LogDet, Hamming and Jaccard distances consistently perform very
poorly. ADD has the best overall performance, producing either the
best results, or results that are not significantly different from
the best results in six of the eight models; it also gives quite reasonable
results in the remaining two cases. PADD has six near-best results,
but fares worse on the remaining two. (The `all' model violates the
assumptions of PADD. Possibly `p\_all' performs poorly because there
are so few markers in each partition.) Log Jaccard is not far behind
the leading methods. DICE, Nei-Li and LogDice round out the middle
of the field. Table~\ref{tab:bad splits Yule} shows rather different
results for the clock-like trees, with the best distances being Jaccard,
DICE, Log Jaccard, PADD then ADD. We were somewhat surprised that
the Jaccard distance did so well here given its poor performance on
the non-clock-like trees. Variation between distance methods is smaller,
as Yule trees have some very short branches, which are hard for any
method to resolve. LogDet performs consistently poorly for both clock-like
and non-clock-like trees.

\begin{figure}[H]
\includegraphics{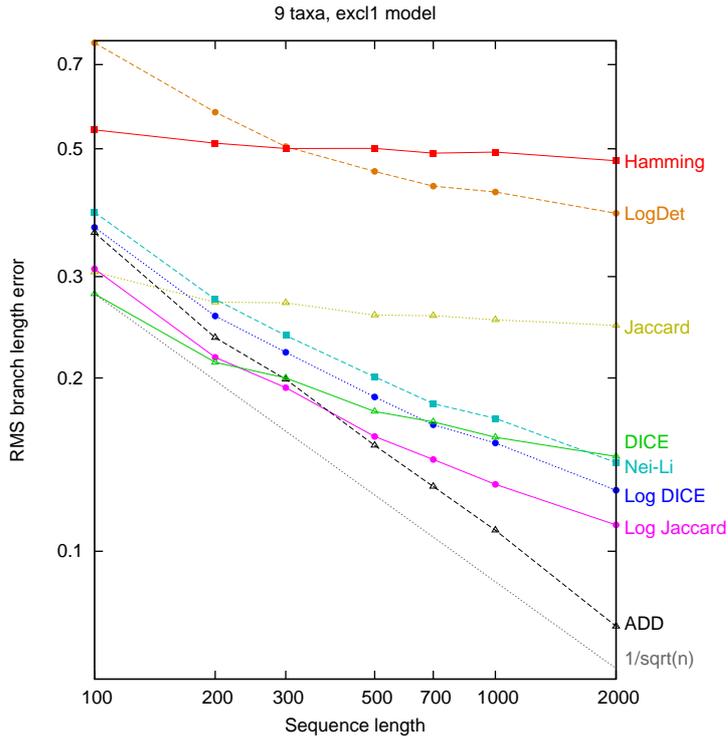}\caption{Root mean square (RMS) error in branch length estimation plotted as
a function of the number of characters for the various distance methods.
If the method is biased, the RMS error cannot fall below the error
caused by the bias, and the line is approximately horizontal (e.g.
Hamming distance.) The grey {}``1/sqrt(n)'' line illustrates the
expected slope of an optimal method, with variance inversely proportional
to sequence length.\label{fig:RMS vs length}}
\end{figure}

Figure~\ref{fig:RMS vs length} plots the accuracy of branch length
reconstruction against sequence length. The Hamming, LogDet, Jaccard
and (to a lesser extent) DICE distances all show signs of ceasing
to improve with increasing sequence length. This is expected when
the method's bias exceeds the sampling error. Only ADD improves at
the optimal rate (lower dotted line.)

\begin{figure}[H]
\includegraphics[width=0.5\textwidth]{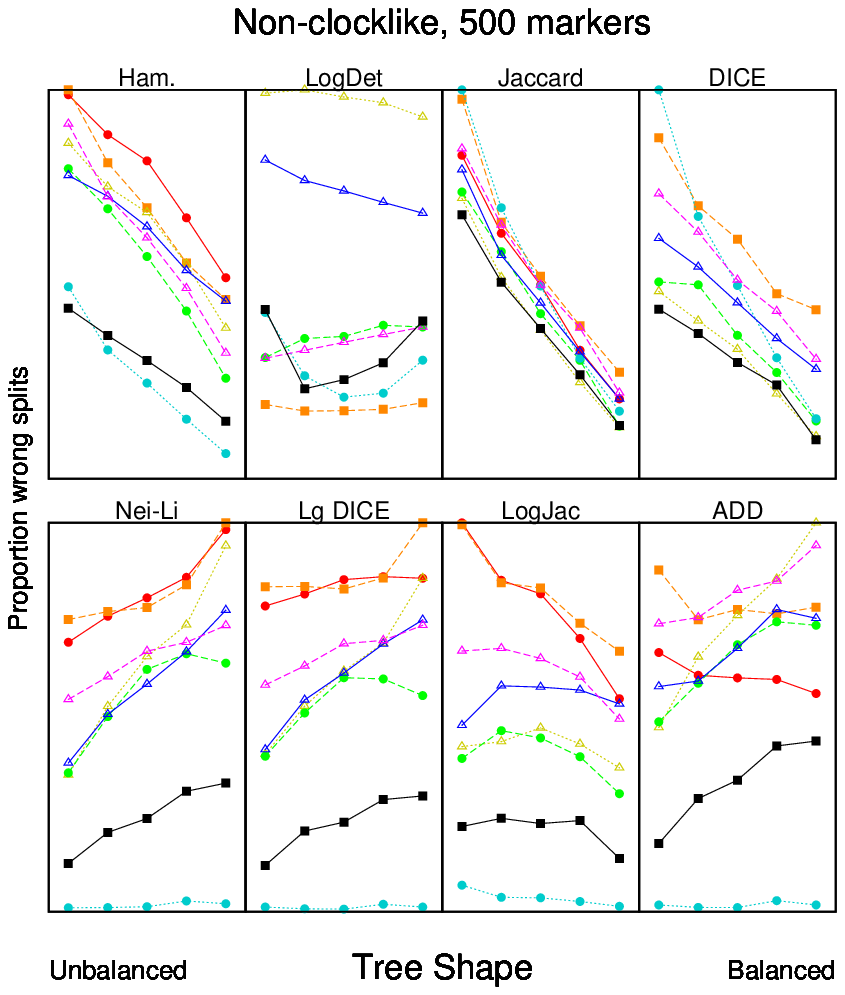}\includegraphics[width=0.5\textwidth]{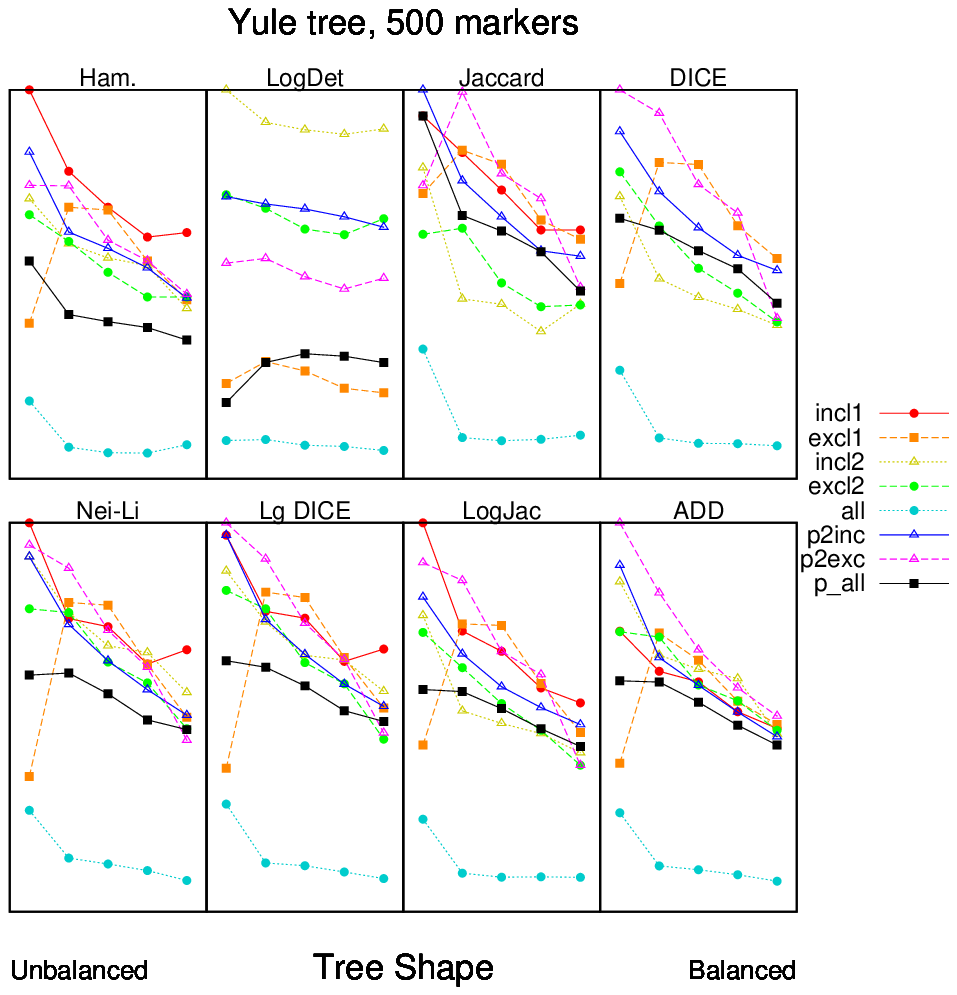}\caption{The effect of tree shape (measured by the number of cherries in the
unrooted true tree) on split error rate. In each panel, unbalanced
trees are on the left, balanced trees on the right. (Points are for
3 to 7 cherries.) Downward sloping lines indicate a bias towards constructing
balanced trees, and upward sloping lines a bias for unbalanced trees.
The vertical axis is linear and originates at zero, but is of different
scale in each panel.\label{fig:effect of cherry number}}
\end{figure}

In figure \ref{fig:effect of cherry number}, we investigate the possibility
of bias in the distances due to tree shape. We divide the true trees
according to how many cherries are in the unrooted tree. (A `cherry'
is an internal node directly connected to exactly two leaves.) The
minimum number is 2 (a maximally unbalanced or `caterpillar' tree).
For 15 taxa, the maximum is 7. Two-cherry trees were too rare to get
reliable statistics, so figure~\ref{fig:effect of cherry number}
shows results for 3 to 7 cherries. If a method is biased towards producing
unbalanced trees, it will be more accurate when the true tree is unbalanced
than when it is balanced. The non-logarithmic methods (Hamming, Jaccard,
DICE) generally have high error rates on unbalanced trees, indicating
a bias in favour of balanced trees. For the other methods, there is
no obvious consistent bias. For example, Nei-Li, LogDice and ADD are
slightly biased towards unbalanced trees for the non-clock like tree
simulations and towards balanced trees for the clock like tree simulations.
More plots demonstrating this lack of consistency are provided in
the supplementary material, figures 1--6.

\subsection*{Applications to real data}

\subsubsection*{Case Study 1 - \emph{ Eucalyptus} DArT data}

We analysed the DArT \emph{ Eucalyptus} dataset of \citet{SteaneEtal}.
This includes 94 species of \emph{ Eucalyptus} from across the full
taxonomic range (excluding \emph{ Corymbia}). The dataset comprised
7490 non-redundant DArT markers (newly acquired sequence data have
allowed us to eliminate 864 markers as redundant, reducing the number
of markers from the 8354 reported by Steane et al. 2011). This dataset
was generated during the development phase of the \emph{ Eucalyptus}
DArT array and only about 32\% of the markers in this dataset are
included on the final, publicly available \emph{ Eucalyptus} DArT
array \citep{SansaloniEtal}. 

\begin{figure}[H]
\includegraphics[width=1\textwidth]{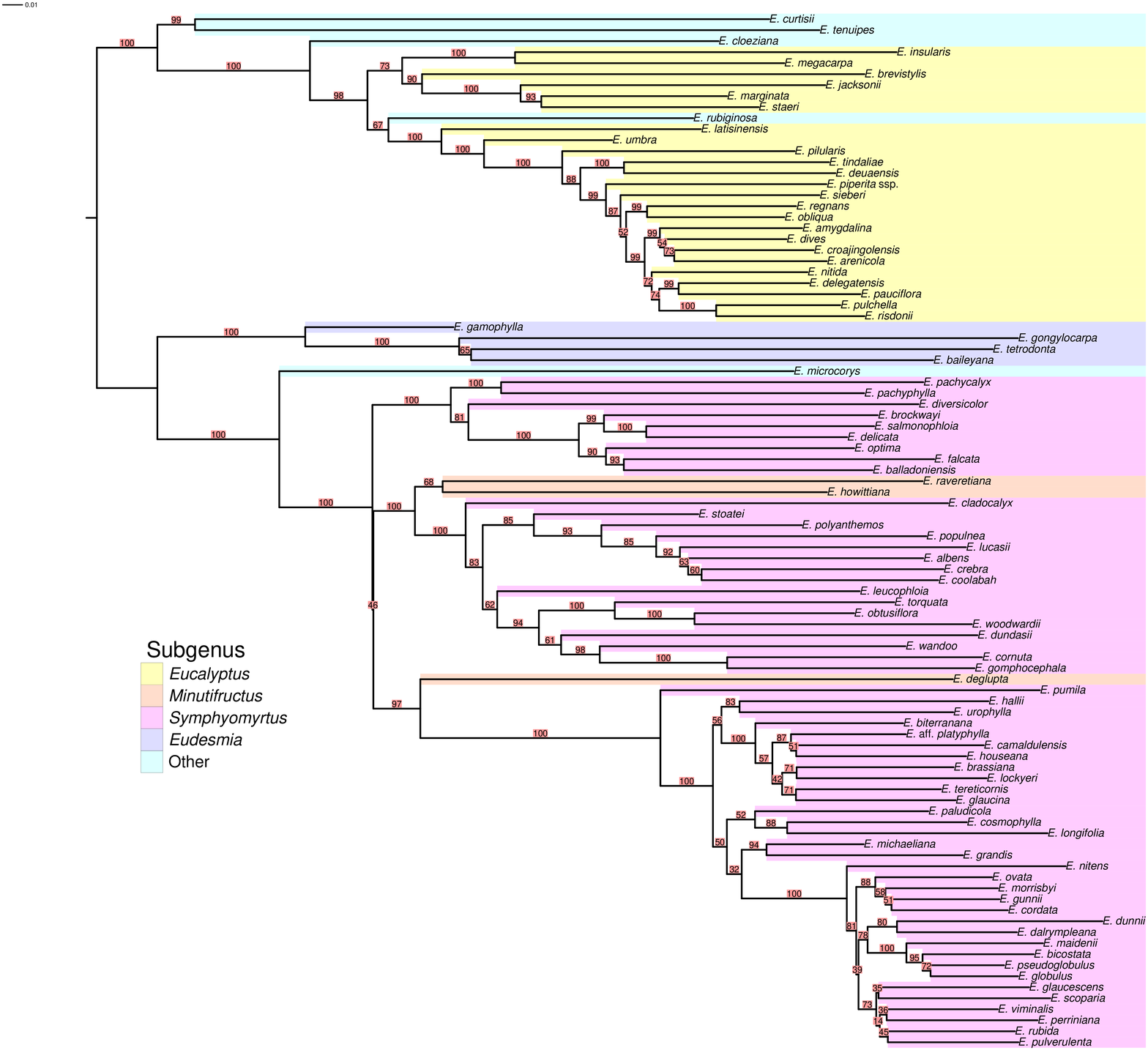}\caption{\emph{Eucalyptus} phylogeny derived from DArT data analyzed with PADD
distance and minimum evolution tree building. \label{fig:Eucalyptus PADD}}
\end{figure}

Analysis of this data using the PADD distance yielded the tree shown
in Figure~\ref{fig:Eucalyptus PADD}. Branch support for this and
subsequent trees are from 1000 nonparametric (resampling randomly
with replacement) bootstraps. Rooting on \emph{E. curtisii} (subgenus
\emph{Acerosae}) was based on previous studies \citep{DrinnanLadiges,SteaneITS}.
The topology of the PADD tree was highly concordant with the most
recently published classification \citep{Brooker} and previous molecular
studies using ITS sequence data \citep{SteaneITS}. The tree was also
highly congruent with a cladistic analysis of the same data \citep{SteaneEtal},
but the PADD tree provided increased resolution at some key nodes.
For example, sections \emph{Latoangulatae} (SL), \emph{Exsertaria}
(SE) and \emph{Racemus} (SR) in the PADD tree form a cluster that
is distinct from all other sections, largely in agreement with results
from ITS sequence data of \citet{SteaneITS}. The cladistic analysis
of the DArT data \citep{SteaneEtal} did not resolve the relationships
between these sections and section \emph{Maidenaria}. However, the
position of section \emph{Racemus} in relation to sections \emph{Latoangulatae}
and \emph{Maidenaria} remains equivocal, with cladistic analysis of
the DArT data and cladistic analysis of ITS sequence data placing
it close to section \emph{Maidenaria}. The bootstrap values on the
branches of the PADD tree are generally high compared to those obtained
by the cladistic analysis. Bootstrap values tended to be higher on
internal nodes where there were longer branches (i.e., splits with
more character support), although bootstrap values were generally
>50\% even when branches were very short. This contrasts with the
cladistic analysis \citep{SteaneEtal} where many branches throughout
the cladogram had <50\% bootstrap support.

\subsubsection*{Case Study 2 - \emph{Adnataria (Eucalyptus)} DArT data}

We screened 90 species of \emph{Eucalyptus} from section \emph{Adnataria}
\citep{Brooker} plus three outgroup taxa (\emph{E. cornuta}, sect.
\emph{Bisectae}; \emph{E. torquata}, sect. \emph{Dumaria}; and \emph{E.
staeri}, sect. \emph{Longistylus}) using the publicly available \emph{Eucalyptus}
DArT array \citep{SansaloniEtal}. Leaf samples were collected from
Currency Creek Arboretum, South Australia; details of the samples
will be given in a subsequent paper, but are available from the authors
upon request. DNA was prepared as described previously \citep{SteaneEtal}.
DArT analysis was conducted by DArT P/L (Canberra, Australia) using
their standard protocol \citep{SansaloniEtal}. 

\begin{figure}[H]
\includegraphics[width=1\textwidth]{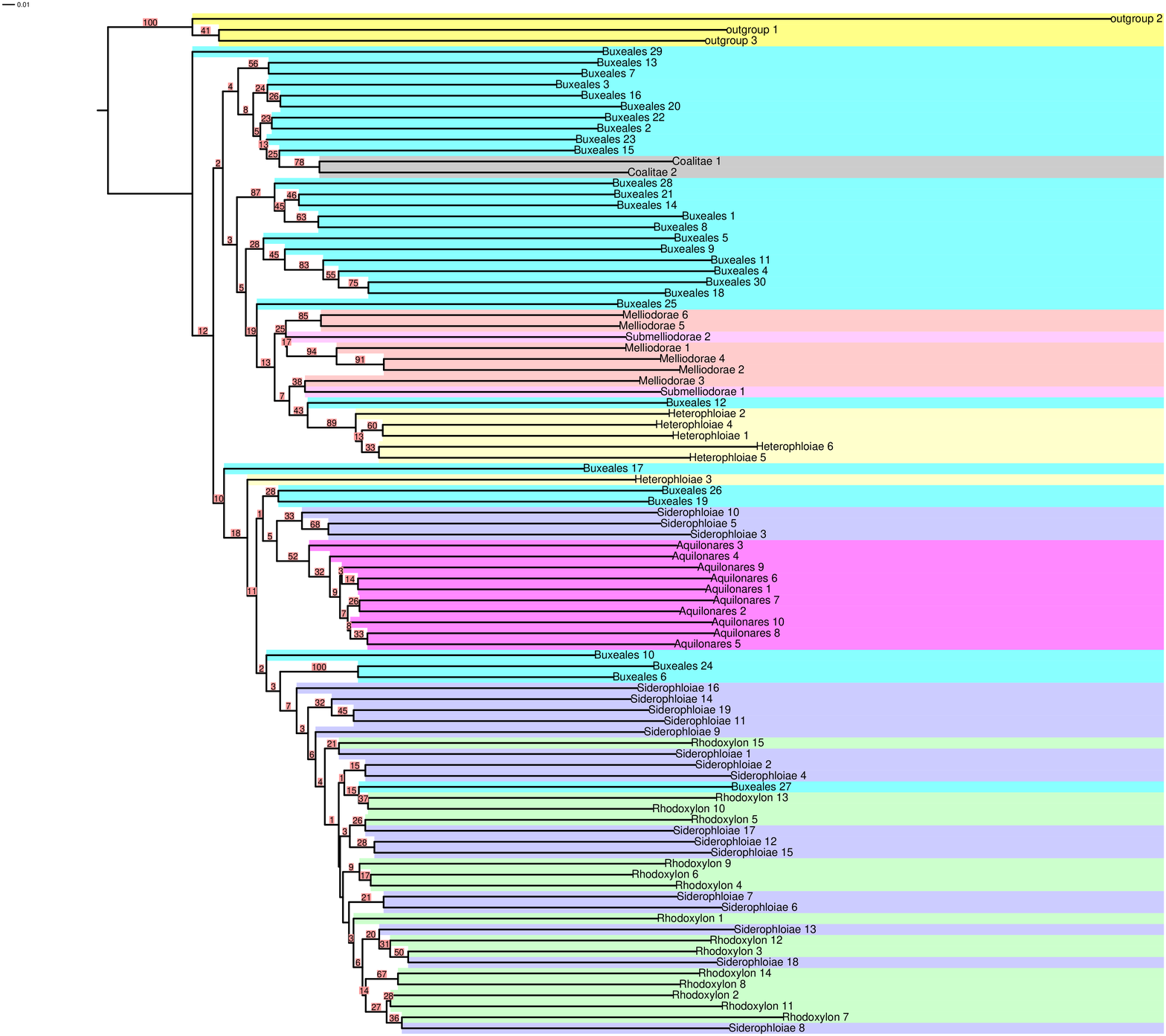}\caption{Phylogeny for \emph{Eucalyptus} section \emph{Adnataria}, derived
from DArT data using PADD distance and minimum evolution tree building.\label{fig:Adnataria PADD}
(Species names anonymized for arXiv submission only.)}
\end{figure}

Section\emph{ Adnataria} (subgenus \emph{Symphyomyrtus}) includes
100--130 terminal taxa of which 90 were included in the DArT analysis.
Because of the large amount of potential homoplasy in DArT data, it
is preferable to include as much genetic variation as possible from
within the study group, in order to minimise the risk of long-branch
attraction and misleading results. Accordingly, the samples in the
study represented eight of the nine series delineated by \citet{Brooker}
and represented the full geographic distribution of the section. (DArT
data for \emph{E. dawsonii}, the single species in series \emph{Dawsonianae},
were not available). Of the 7680 markers on the DArT array, 3707 provided
potentially phylogenetically informative data, of which 1230 were
later found to be redundant and removed from the analysis, leaving
2477 markers to be analysed. The topology of the PADD tree shown in
figure~\ref{fig:Adnataria PADD} is not entirely congruent with the
established classification \citep{Brooker}, but it does have some
interesting features. While at first glance it appears that - apart
from series \emph{Aquilonares} - most of the series within section
\emph{Adnataria} are polyphyletic, on closer inspection the series
cluster into more-or-less discrete groups. Series \emph{Rhodoxylon}
and \emph{Siderophloiae} form a distinct cluster (apart from the intrusion
of\emph{ E. rummeryi}, series \emph{Buxeales}) even though these series
are in different subsections \citep{Brooker}. Most of series \emph{Buxeales}
and supraspecies \emph{Mollucanae} form a discrete cluster that is
{}``sister\textquotedblright{} to a cluster comprising series \emph{Heterophloiae},
\emph{Melliodorae} and \emph{Submelliodorae} (and one species from
series\emph{ Buxeales}), none of which are considered by \citet{Brooker}
to be particularly close. Close examination of the data reveals that
there may be a biogeographic aspect to the clusters identified by
the PADD algorithm. This is being explored in a separate paper (Steane
et al., in prep.).

Notable features of figure~\ref{fig:Adnataria PADD} are the shortness
of the internal branches and the poor bootstrap values. The internal
edges of this tree have mean bootstrap support of 29\% and ratio of
internal to total edge lengths (\textquotedbl{}stemminess\textquotedbl{},
Fiala and Sokal, 1985) of 0.052. Compared to the \emph{Adnatari}a
phylogeny (figure~\ref{fig:Adnataria PADD}), the \emph{Eucalyptus}
phylogeny has much higher bootstrap supports (mean 81\%) and higher
stemminess (0.138). The \emph{Adnataria} dataset differs from \emph{Eucalyptus}
in that it has fewer markers (2477 compared to 7490) and its taxa
are much more closely related to one another (i.e., they span a much
narrower taxonomic range). To test whether the structural and statistical
differences between the trees were simply a function of sample size
(i.e., the number of markers), we generated 100 samples of 2477 markers
from the \emph{Eucalyptu}s dataset (randomly chosen with replacement)
and then bootstrapped each sample 1000 times. The resampled \emph{Eucalyptus}
data had mean bootstrap support of 60\% and mean stemminess of 0.153
(std. deviation 0.003), demonstrating that the differences are not
simply due to sample size. We conclude that the short internal branch
lengths in the \emph{Adnataria} phylogeny may be indicative of the
aftermath of a period of adaptive radiation that produced many species
over a short period of time, or of a loss of phylogenetic signal as
a result of hybridization (or both).

If a marker is monomorphic (all 0 or all 1) within a dataset, the
DArT data-processing cannot distinguish whether it is present or absent
\citep[fig. 4 of][]{DArT_intro}. Monomorphic data are automatically
excluded from the final dataset. This is a potential problem with
the\emph{ Adnataria} dataset, as it is much less genetically diverse
than the full \emph{Eucalyptus} genus from which the array was developed.
We are not concerned by markers which are absent from all samples
being omitted from the data, as they are ignored by the ADD formula,
but there is a prospect that some markers which are present in all
taxa have been omitted. To test the effect of such an omission, we
added to the dataset 500 markers that were scored as present in all
taxa, and reanalysed. The distances were uniformly smaller, there
was very little change to the topology, bootstrap values or stemminess.
We conclude that this potential for missing data does not materially
affect our results. (The trees for this test are shown in the supplementary
material, figures 7 and 8.)

\subsubsection*{Case study 3 - Gene Family Bacterial Phylogeny}

\begin{figure}[H]
\includegraphics[width=1\textwidth]{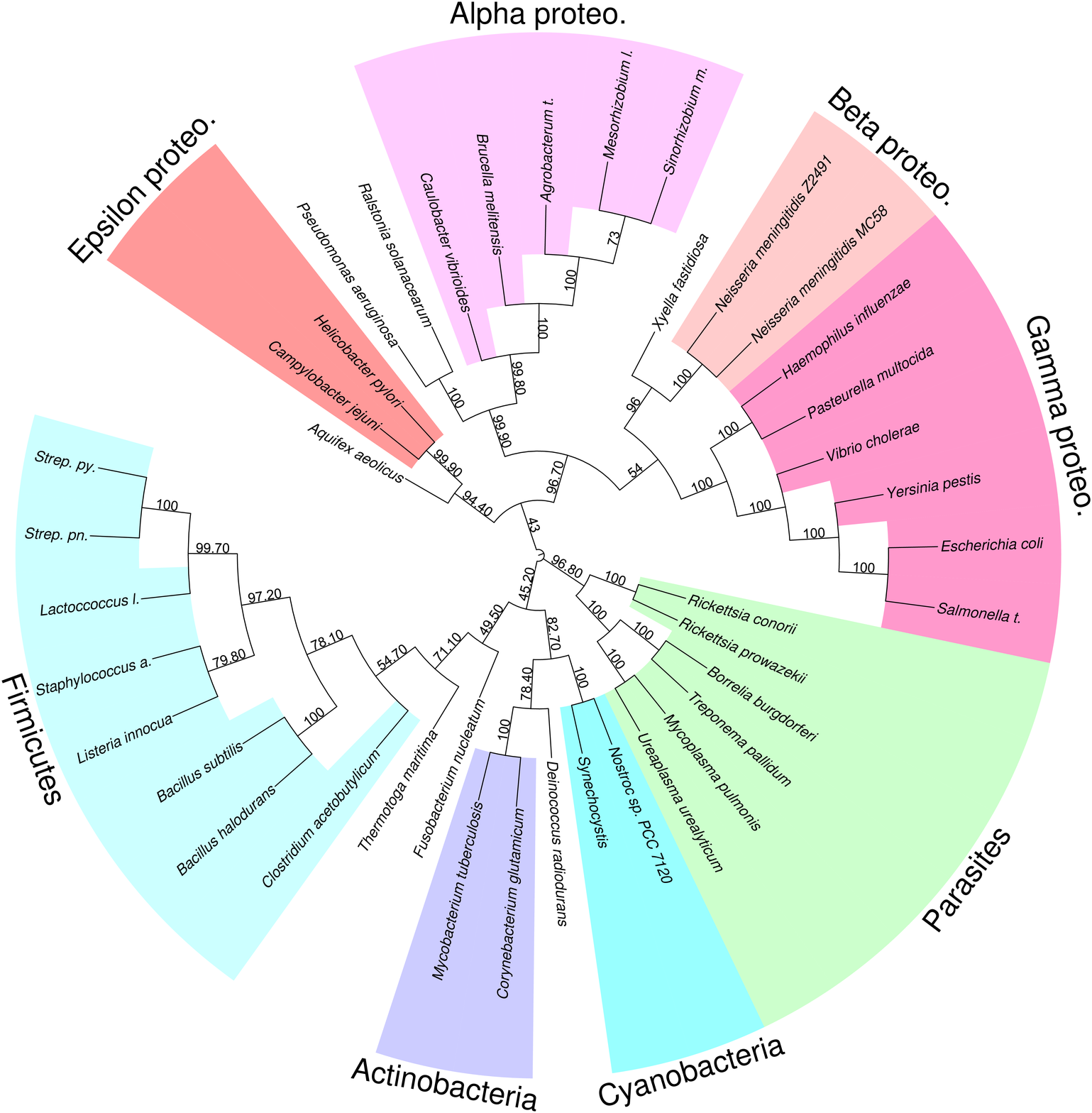}\caption{Gene family presence/absence phylogeny with bootstrap support from
\citet[ figure 9]{SBS}. (Edges are not drawn to scale.) \label{fig:SBS tree}}
\end{figure}

In this case study we reanalyse the gene presence/absence data used
by \citet{SBS}. A potential weakness of the conditioned genome reconstruction
approach which has attracted attention recently is the influence of
the choice of conditioning genome on the outcome. The most sophisticated
attempt to address this, \citet{SBS}, used many conditioning genomes,
produced a tree for each one, and finally constructed a consensus
tree from these. The tree they derived for 40 of the bacterial genomes
is shown in figure \ref{fig:SBS tree}. 

It has been noted previously \citep{LakeRivera,SBS} that parallel
loss of genes required for free living causes parasitic bacteria from
all bacterial phyla to falsely group together in this analysis.

\begin{figure}[H]
\includegraphics[width=1\columnwidth]{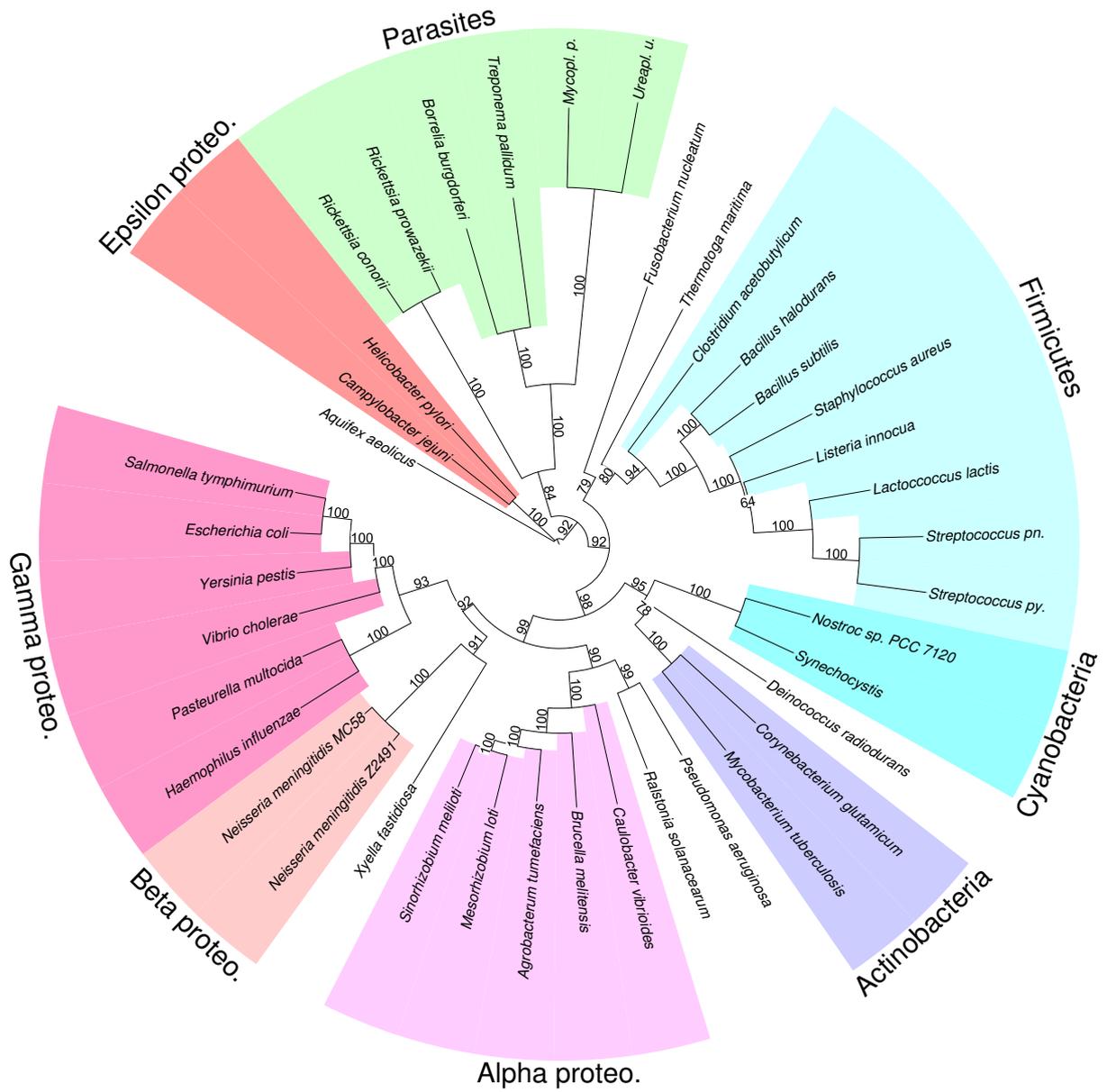}\caption{Phylogeny of bacteria by COG gene family presence/absence data, using
the additive Dollo distance. Edges are to scale. \label{fig:COG ADD phylogeny}}
\end{figure}

As we have derived ADD as a special case of LogDet that is applicable
to a Dollo process, we can analyze these COG data without the need
for a conditioning genome. In effect, where \citet{LakeRivera} use
LogDet distances and a conditioning genome to find $n_{00}$, we use
LogDet distances assuming $n_{00}$ is infinite (equation~\ref{eq:limit log det}.)
The resulting phylogeny is shown in figure~\ref{fig:COG ADD phylogeny}.
It differs from the \citet{SBS} phylogeny on only three edges, two
of which have less than 50\% bootstrap support in their tree.

\section*{Discussion}

ADD is a simple, consistent distance suitable for use with binary
data which evolve under a stochastic Dollo process. We have shown
that ADD is also consistent for data which has been censored in such
a way that only traits that are present in a single reference taxon
are observable. Censoring by multiple reference taxa is more complex
but in the case where each trait is known to derive from a particular
reference taxon (e.g. in the DArT data presented above), ADD can be
extended by a straight-forward partitioning scheme (PADD).

Our simulation supports the expectation from theory that ADD and PADD
should perform well on data generated under a stochastic Dollo model
with various censoring schemes. By contrast, several other distances
that are in common use for both DArT data and for analysis of gene-content
data do not perform well in the simulations. This suggests that they
should probably not be used for data which are thought to be generated
under a Dollo model. We used DArT data and gene presence/absence data
as illustrations of our new approach, but our distances can be applied
to any data derived from complex traits that are unlikely to evolve
more than once independently.

More specifically, our simulations indicate that LogDet performs very
poorly on stochastic Dollo data. This is a concern for the use of
LogDet distances in conditioned genome reconstruction \citep{LakeRivera,SBS}.
The use of LogDet on gene presence/absence data in deep bacterial
phylogeny not only necessitates an extra layer of complication with
conditioning genomes to find the number of shared absences, but also
depends critically on horizontal gene transfer (HGT) being sufficiently
common to justify the use of a Markov model. For the gene-content
phylogeny based on the COG data, it is interesting that the tree found
using ADD is very similar to the tree found by \citet{SBS}. Indeed
the two approaches have very different underlying assumptions. ADD
ignores the possibility of HGT, which certainly does occur for these
data, and the augmented conditioning genome reconstruction method
of \citet{SBS} ignores the Dollo aspect of the data treating shared
absences and shared presences equivalently. Both methods recover the
(presumed artefactual) parasite clade \citep{Sangaralingam}. We could
imagine an intermediate model for this type of data, where markers
primarily evolve via a stochastic Dollo process, but are occasionally
subject to horizontal gene transfer. In this situation, a suitable
distance might be the LogDet distance, with $n_{00}$ augmented, but
still finite.

ADD also appears to be an appropriate distance to use for DArT data,
and we have applied it to two closely related datasets. The results
obtained using the PADD algorithm for the \emph{Eucalyptus} data are
satisfyingly congruent with traditional morphology-based taxonomies
(e.g. \citealp{Brooker}) and phylogenies based on ITS sequence data.
In addition, the PADD analysis of DArT data appears to somewhat improve
on the parsimony-based analysis as it provides more resolution, possibly
because the parsimony-based approach does not take account of branch
length information and can thus be more easily misled by homoplasy.

In pilot studies we demonstrated that DArT data had the potential
to resolve relationships among closely related \emph{Eucalyptus} species
\citep{SteaneEtal}. This level of resolution has always been elusive
to eucalypt systematists for various reasons including recent/incomplete
speciation and the high incidence of inter-specific hybridisation
\citep{Bryne}. To test the efficacy of the partitioned additive Dollo
distance at this level of divergence we applied it to a set of DArT
data for section \emph{Adnataria}. The PADD-derived \emph{Adnataria}
phylogeny suggests that the series delimited by \citet{Brooker} are
not robust groups. However, the lack of bootstrap support and the
short branch lengths do not provide us with confidence that PADD analysis
of the DArT data has uncovered the {}``true tree''. Further analyses
of this dataset (currently underway) may reveal patterns of variation
in other traits (e.g., morphology, physiology, biogeography) that
will inform us about the plausibility of this DArT-based phylogeny
derived using the PADD algorithm.

With the increasing debate about the appropriateness of the Tree of
Life metaphor for many domains of life, it would be timely to attempt
to extend the basic Dollo model to incorporate borrowing of traits.
Indeed, both the bacterial gene-content example and the \emph{Adnataria}
example are cases where Dollo models that include {}``borrowing''
of traits would be very appropriate. Another avenue to pursue would
be the development of consistent distances in the case of multi-state
Dollo models (such as discussed by \citealp{StochasticDollo1}). With
the ever-increasing abundance of genomic data, finding good models
for the evolution of complex traits is as appropriate now as it was
in the time of Dollo.

\section*{Acknowledgments}

This research was funded by BH's ARC Future Fellowship FT100100031
and ARC grant DP0986491 to Brad Potts. This manuscript was completed
while DS was a CRN (Collaborative Research Network) research fellow
at the University of the Sunshine Coast. VM thanks the Royal Society
and the New Zealand Marsden Fund for supporting his visit to New Zealand
where work started on this project. Andrzej Kilian helped us understand
details of the DArT data processing pipeline. Lars Jermin suggested
considering the triangle inequality and additivity properties of our
distances. We also acknowledge Mark Robertson for assistance in the
field, Sascha Wise for technical laboratory assistance, Ren\'e Vaillancourt
and Brad Potts for academic input, and the Central Science Laboratory
at the University of Tasmania for use of their facilites.

\bibliographystyle{sysbio}
\bibliography{add}

\section*{\newpage{}Tables}

\begin{table}[H]
\begin{centering}
\begin{tabular}{cccccc}
\hline 
{\footnotesize Distance}  & {\footnotesize DICE}  & {\footnotesize LogJaccard}  & {\footnotesize LogDICE}  & {\footnotesize LogDet}  & {\footnotesize ADD}\tabularnewline
\hline 
 & {\footnotesize A:10}  & {\footnotesize A:00111}  & {\footnotesize A:001}  & {\footnotesize A:00101}  & {\footnotesize A:011}\tabularnewline
{\footnotesize Alignment:}  & {\footnotesize B:01}  & {\footnotesize B:11001}  & {\footnotesize B:111}  & {\footnotesize B:00011}  & {\footnotesize B:101}\tabularnewline
 & {\footnotesize C:11}  & {\footnotesize C:11111}  & {\footnotesize C:011}  & {\footnotesize C:00001}  & {\footnotesize C:111}\tabularnewline
\hline 
{\footnotesize $d(A,B)$}  & {\footnotesize 1}  & {\footnotesize log(5)=1.61}  & {\footnotesize log(2)=0.69}  & {\footnotesize log(6)=1.79}  & {\footnotesize log(4)=1.39}\tabularnewline
{\footnotesize $d(A,C)+d(B,C)$}  & {\footnotesize 2/3}  & {\footnotesize 2 log(5/3)=1.02}  & {\footnotesize log(1.5)+log(1.25)=0.63}  & {\footnotesize log(8/3)=0.98}  & {\footnotesize 2 log(1.5)=0.81}\tabularnewline
\hline 
\end{tabular}
\par\end{centering}

\caption{Counterexamples to the Triangle Inequality\label{tab:Triangle inequality}}
\end{table}
\begin{table}[H]
\begin{centering}
\begin{tabular}{ccccccccc}
\hline 
{\small Distance}  & {\small Hamming}  & {\small Jaccard}  & {\small DICE}  & {\small LogJaccard}  & {\small LogDICE}  & {\small Nei-Li}  & {\small LogDet}  & {\small ADD}\tabularnewline
\hline 
{\small Triangle ineq.}  & {\small Yes$^{a}$}  & {\small Yes$^{b}$}  & {\small No}  & {\small No}  & {\small No}  & {\small Unknown}  & {\small No}  & {\small No}\tabularnewline
{\small Additive}  & {\small No}  & {\small No}  & {\small No}  & {\small Unknown}  & {\small Unknown}  & {\small Unknown}  & {\small Yes$^{c}$}  & {\small Yes}\tabularnewline
\hline 
\end{tabular}
\par\end{centering}

\emph{a}: \citealt{Hamming}; \emph{b}: \citealt{JaccardTEQ}; \emph{c}:
\citealt{paralinear}

\caption{Summary of the mathematical properties of the distances tested in
this paper.\label{tab:properties of distances}}
\end{table}
\begin{table}[H]
\begin{tabular}{cllllllll}
\hline 
{\scriptsize Method} & {\scriptsize incl1} & {\scriptsize excl1} & {\scriptsize incl2} & {\scriptsize excl2} & {\scriptsize all} & {\scriptsize p2inc} & {\scriptsize p2exc} & {\scriptsize p\_all}\tabularnewline
\hline 
\hline 
{\scriptsize LogDet} & {\scriptsize n/a} & {\scriptsize \cellcolor[rgb]{1,.5,0.}0.0624(15)} & {\scriptsize \cellcolor[rgb]{1,0,0.}0.3472(227)} & {\scriptsize \cellcolor[rgb]{1,0,0.}0.1313(64)} & {\scriptsize \cellcolor[rgb]{1,0,0.}0.0824(273)} & {\scriptsize \cellcolor[rgb]{1,0,0.}0.2616(121)} & {\scriptsize \cellcolor[rgb]{1,0,0.}0.1242(42)} & {\scriptsize \cellcolor[rgb]{1,0,0.}0.0950(63)}\tabularnewline
\hline 
{\scriptsize Hamming} & {\scriptsize \cellcolor[rgb]{1.,0.,0.}0.1204(51)} & {\scriptsize \cellcolor[rgb]{1,0,0.}0.1047(35)} & {\scriptsize \cellcolor[rgb]{1,0,0.}0.1008(59)} & {\scriptsize \cellcolor[rgb]{1,0,0.}0.0857(37)} & {\scriptsize \cellcolor[rgb]{1,0,0.}0.0370(121)} & {\scriptsize \cellcolor[rgb]{1,0,0.}0.0966(36)} & {\scriptsize \cellcolor[rgb]{1,0,0.}0.0931(28)} & {\scriptsize \cellcolor[rgb]{1,0,0.}0.0457(26)}\tabularnewline
\hline 
{\scriptsize Jaccard} & {\scriptsize \cellcolor[rgb]{1.,0,0.}0.0649(22)} & {\scriptsize \cellcolor[rgb]{1,.15,0.}0.0699(19)} & {\scriptsize \cellcolor[rgb]{1,0,0.}0.0508(25)} & {\scriptsize \cellcolor[rgb]{1,0,0.}0.0578(21)} & {\scriptsize \cellcolor[rgb]{1,0,0.}0.0665(220)} & {\scriptsize \cellcolor[rgb]{1,0.25,0.}0.0600(17)} & {\scriptsize \cellcolor[rgb]{1,0.38,0.}0.0679(16)} & {\scriptsize \cellcolor[rgb]{1,0,0.}0.0512(30)}\tabularnewline
\hline 
{\scriptsize DICE} & {\scriptsize \cellcolor[rgb]{1,.96,0.}0.0435(10)} & {\scriptsize \cellcolor[rgb]{0.74,1,0.}0.0464(7)} & {\scriptsize \cellcolor[rgb]{.69,1,0.}0.0246(7)} & {\scriptsize \cellcolor[rgb]{0.47,1,0.}0.0288(5)} & {\scriptsize \cellcolor[rgb]{1,0,0.}0.0381(125)} & {\scriptsize \cellcolor[rgb]{0.45,1,0.}0.0347(4)} & {\scriptsize \cellcolor[rgb]{0.35,1,0.}0.0402(3)} & {\scriptsize \cellcolor[rgb]{0.86,1,0.}0.0231(9)}\tabularnewline
\hline 
{\scriptsize Nei-Li} & {\scriptsize \cellcolor[rgb]{1,.36,0.}0.0549(16)} & {\scriptsize \cellcolor[rgb]{1,0.90,0.}0.0541(11)} & {\scriptsize \cellcolor[rgb]{1,.01,0.}0.0438(20)} & {\scriptsize \cellcolor[rgb]{1,0.86,0.}0.0405(11)} & {\scriptsize \cellcolor[rgb]{0.17,1,0.}}\textbf{\scriptsize 0.0011}{\scriptsize (2)} & {\scriptsize \cellcolor[rgb]{0.71,1,0.}0.0398(7)} & {\scriptsize \cellcolor[rgb]{0.55,1,0.}0.0446(6)} & {\scriptsize \cellcolor[rgb]{0.39,1,0.}0.0168(4)}\tabularnewline
\hline 
{\scriptsize LogDICE} & {\scriptsize \cellcolor[rgb]{1,.63,0.}0.0497(14)} & {\scriptsize \cellcolor[rgb]{0.89,1,0.}0.0495(9)} & {\scriptsize \cellcolor[rgb]{1,.52,0.}0.0363(15)} & {\scriptsize \cellcolor[rgb]{0.75,1,0.}0.0337(8)} & {\scriptsize \cellcolor[rgb]{0,1,0.}}\textbf{\scriptsize 0.0006} & {\scriptsize \cellcolor[rgb]{0.52,1,0.}0.0362(5)} & {\scriptsize \cellcolor[rgb]{0.33,1,0.}0.0398(3)} & {\scriptsize \cellcolor[rgb]{0.17,1,0.}}\textbf{\scriptsize 0.0140}{\scriptsize (2)}\tabularnewline
\hline 
{\scriptsize LogJaccard} & {\scriptsize \cellcolor[rgb]{0.88,1,0.}0.0406(9)} & {\scriptsize \cellcolor[rgb]{0.51,1,0.}0.0416(5)} & {\scriptsize \cellcolor[rgb]{.58,1,0.}0.0230(6)} & {\scriptsize \cellcolor[rgb]{0.09,1,0.}}\textbf{\scriptsize 0.0222}{\scriptsize (1)} & {\scriptsize \cellcolor[rgb]{0.38,1,0.}0.0017(4)} & {\scriptsize \cellcolor[rgb]{0.17,1,0.}}\textbf{\scriptsize 0.0293}{\scriptsize (2)} & {\scriptsize \cellcolor[rgb]{0.01,1,0.}}\textbf{\scriptsize 0.0329}{\scriptsize (0)} & {\scriptsize \cellcolor[rgb]{0,1,0.}}\textbf{\scriptsize 0.0117}\tabularnewline
\hline 
{\scriptsize ADD} & {\scriptsize \cellcolor[rgb]{0,1,0.}}\textbf{\scriptsize 0.0241} & {\scriptsize \cellcolor[rgb]{0,1,0.}}\textbf{\scriptsize 0.0308} & {\scriptsize \cellcolor[rgb]{1,.92,0.}0.0303(11)} & {\scriptsize \cellcolor[rgb]{0.36,1,0.}0.0269(4)} & {\scriptsize \cellcolor[rgb]{0,1,0.}}\textbf{\scriptsize 0.0006} & {\scriptsize \cellcolor[rgb]{0.06,1,0.}}\textbf{\scriptsize 0.0273}{\scriptsize (1)} & {\scriptsize \cellcolor[rgb]{0,1,0.}}\textbf{\scriptsize 0.0326} & {\scriptsize \cellcolor[rgb]{0.16,1,0.}}\textbf{\scriptsize 0.0139}{\scriptsize (2)}\tabularnewline
\hline 
{\scriptsize PADD} & {\scriptsize \cellcolor[rgb]{0,1,0.}}\textbf{\scriptsize 0.0241} & {\scriptsize \cellcolor[rgb]{0,1,0.}}\textbf{\scriptsize 0.0308} & {\scriptsize \cellcolor[rgb]{0,1,0.}}\textbf{\scriptsize 0.0146} & {\scriptsize \cellcolor[rgb]{0,1,0.}}\textbf{\scriptsize 0.0206} & {\scriptsize \cellcolor[rgb]{1,0,0.}0.0082(25)} & {\scriptsize \cellcolor[rgb]{0,1,0.}}\textbf{\scriptsize 0.0260} & {\scriptsize \cellcolor[rgb]{0.13,1,0.}}\textbf{\scriptsize 0.0354}{\scriptsize (1)} & {\scriptsize \cellcolor[rgb]{1,0.31,0.}0.0341(17)}\tabularnewline
\hline 
\hline 
{\scriptsize DolloP} & {\scriptsize 0.0396} & {\scriptsize 0.0363} & {\scriptsize 0.0236} & {\scriptsize 0.0228} & {\scriptsize 0.0004} & {\scriptsize 0.0245} & {\scriptsize 0.0296} & {\scriptsize 0.0074}\tabularnewline
\hline 
\end{tabular}

\caption{The proportion of incorrect splits in minimum evolution trees derived
from the various distances. Columns show the results for each of the
various models of data censoring. Numbers in parentheses indicate
how many standard deviations worse this result is than the best result
for this model. Within each column, for the distance-based methods,
the best (lowest) value is shown in bold, along with any which are
not significantly worse than the best value. Distance-based results
are colour coded, from green (best in column) to red (at least 20
standard deviations worse than the best.) Also included for comparison
are the Dollo parsimony results. These data are derived from simulations
using 15 taxa and a mean of 500 markers per taxon, with 5000 random
trials.\label{tab:bad splits}}
\end{table}
\begin{table}[H]
\begin{tabular}{cllllllll}
\hline 
{\scriptsize Method} & {\scriptsize incl1} & {\scriptsize excl1} & {\scriptsize incl2} & {\scriptsize excl2} & {\scriptsize all} & {\scriptsize p2inc} & {\scriptsize p2exc} & {\scriptsize p\_all}\tabularnewline
\hline 
\hline 
{\scriptsize LogDet} & {\scriptsize n/a} & {\scriptsize \cellcolor[rgb]{1,0.09,0.}0.0677(19)} & {\scriptsize \cellcolor[rgb]{1,0,0.}0.2309(139)} & {\scriptsize \cellcolor[rgb]{1,0,0.}0.1663(88)} & {\scriptsize \cellcolor[rgb]{1,0,0.}0.0223(24)} & {\scriptsize \cellcolor[rgb]{1,0,0.}0.1769(80)} & {\scriptsize \cellcolor[rgb]{1,0,0.}0.1326(50)} & {\scriptsize \cellcolor[rgb]{1,0,0.}0.0811(31)}\tabularnewline
\hline 
{\scriptsize Hamming} & {\scriptsize \cellcolor[rgb]{0.80,1,0.}0.0432(8)} & {\scriptsize \cellcolor[rgb]{0.56,1,0.}0.0401(6)} & {\scriptsize \cellcolor[rgb]{1,0.76,0.}0.0355(12)} & {\scriptsize \cellcolor[rgb]{0.82,1,0.}0.0327(8)} & {\scriptsize \cellcolor[rgb]{0.07,1,0.}}\textbf{\scriptsize 0.0044}{\scriptsize (1)} & {\scriptsize \cellcolor[rgb]{.61,1,0.}0.0365(6)} & {\scriptsize \cellcolor[rgb]{0.43,1,0.}0.0385(4)} & {\scriptsize \cellcolor[rgb]{0.10,1,0.}}\textbf{\scriptsize 0.0253}{\scriptsize (1)}\tabularnewline
\hline 
{\scriptsize Jaccard} & \textbf{\scriptsize \cellcolor[rgb]{0,1,0.}0.0272} & \textbf{\scriptsize \cellcolor[rgb]{0,1,0.}0.0287} & \textbf{\scriptsize \cellcolor[rgb]{0,1,0.}0.0162} & \textbf{\scriptsize \cellcolor[rgb]{0,1,0.}0.0189} & \textbf{\scriptsize \cellcolor[rgb]{0,1,0.}0.0039} & \textbf{\scriptsize \cellcolor[rgb]{0,1,0.}0.0248} & \textbf{\scriptsize \cellcolor[rgb]{0,1,0.}0.0295} & \textbf{\scriptsize \cellcolor[rgb]{0,1,0.}0.0235}\tabularnewline
\hline 
{\scriptsize DICE} & {\scriptsize \cellcolor[rgb]{0.27,1,0.}0.0325(3)} & {\scriptsize \cellcolor[rgb]{0.24,1,0.}}\textbf{\scriptsize 0.0336}{\scriptsize (2)} & {\scriptsize \cellcolor[rgb]{.30,1,0.}0.0209(3)} & {\scriptsize \cellcolor[rgb]{0.30,1,0.}0.0240(3)} & {\scriptsize \cellcolor[rgb]{0.05,1,0.}}\textbf{\scriptsize 0.0042}{\scriptsize (0)} & {\scriptsize \cellcolor[rgb]{0.20,1,0.}}\textbf{\scriptsize 0.0285}{\scriptsize (2)} & {\scriptsize \cellcolor[rgb]{0.20,1,0.}}\textbf{\scriptsize 0.0337}{\scriptsize (2)} & {\scriptsize \cellcolor[rgb]{0.13,1,0.}}\textbf{\scriptsize 0.0259}{\scriptsize (1)}\tabularnewline
\hline 
{\scriptsize Nei-Li} & {\scriptsize \cellcolor[rgb]{0.68,1,0.}0.0408(7)} & {\scriptsize \cellcolor[rgb]{0.65,1,0.}0.0420(7)} & {\scriptsize \cellcolor[rgb]{1,.47,0.}0.0399(15)} & {\scriptsize \cellcolor[rgb]{1,0.93,0.}0.0369(11)} & {\scriptsize \cellcolor[rgb]{0.39,1,0.}0.0069(4)} & {\scriptsize \cellcolor[rgb]{0.62,1,0.}0.0365(6)} & {\scriptsize \cellcolor[rgb]{0.54,1,0.}0.0408(5)} & {\scriptsize \cellcolor[rgb]{0.43,1,0.}0.0314(4)}\tabularnewline
\hline 
{\scriptsize LogDICE} & {\scriptsize \cellcolor[rgb]{.61,1,0.}0.0393(6)} & {\scriptsize \cellcolor[rgb]{0.59,1,0.}0.0407(6)} & {\scriptsize \cellcolor[rgb]{1,.70,0.}0.0364(13)} & {\scriptsize \cellcolor[rgb]{0.95,1,0.}0.0347(9)} & {\scriptsize \cellcolor[rgb]{0.31,1,0.}0.0062(3)} & {\scriptsize \cellcolor[rgb]{0.55,1,0.}0.0353(6)} & {\scriptsize \cellcolor[rgb]{0.49,1,0.}0.0395(5)} & {\scriptsize \cellcolor[rgb]{0.39,1,0.}0.0308(4)}\tabularnewline
\hline 
{\scriptsize LogJaccard} & {\scriptsize \cellcolor[rgb]{0.34,1,0.}0.0340(3)} & {\scriptsize \cellcolor[rgb]{0.34,1,0.}0.0357(3)} & {\scriptsize \cellcolor[rgb]{.59,1,0.}0.0254(6)} & {\scriptsize \cellcolor[rgb]{0.51,1,0.}0.0275(5)} & {\scriptsize \cellcolor[rgb]{0.13,1,0.}}\textbf{\scriptsize 0.0049(1)} & {\scriptsize \cellcolor[rgb]{0.28,1,0.}0.0302(3)} & {\scriptsize \cellcolor[rgb]{0.27,1,0.}0.0352(3)} & {\scriptsize \cellcolor[rgb]{0.18,1,0.}}\textbf{\scriptsize 0.0269}{\scriptsize (2)}\tabularnewline
\hline 
{\scriptsize ADD} & {\scriptsize \cellcolor[rgb]{0.32,1,0.}0.0336(3)} & {\scriptsize \cellcolor[rgb]{0.38,1,0.}0.0364(4)} & {\scriptsize \cellcolor[rgb]{1,.68,0.}0.0366(13)} & {\scriptsize \cellcolor[rgb]{0.94,1,0.}0.0346(9)} & {\scriptsize \cellcolor[rgb]{0.31,1,0.}0.0062(3)} & {\scriptsize \cellcolor[rgb]{0.46,1,0.}0.0336(5)} & {\scriptsize \cellcolor[rgb]{0.46,1,0.}0.0390(5)} & {\scriptsize \cellcolor[rgb]{0.41,1,0.}0.0310(4)}\tabularnewline
\hline 
{\scriptsize PADD} & {\scriptsize \cellcolor[rgb]{0.32,1,0.}0.0336(3)} & {\scriptsize \cellcolor[rgb]{0.38,1,0.}0.0364(4)} & {\scriptsize \cellcolor[rgb]{0.48,1,0.}0.0237(5)} & {\scriptsize \cellcolor[rgb]{0.57,1,0.}0.0284(6)} & {\scriptsize \cellcolor[rgb]{0.90,1,0.}0.0108(9)} & {\scriptsize \cellcolor[rgb]{0.42,1,0.}0.0238(4)} & {\scriptsize \cellcolor[rgb]{0.49,1,0.}0.0396(5)} & {\scriptsize \cellcolor[rgb]{0.70,1,0.}0.0364(7)}\tabularnewline
\hline 
\hline 
{\scriptsize DolloP} & {\scriptsize 0.0411} & {\scriptsize 0.0404} & {\scriptsize 0.0288} & {\scriptsize 0.0311} & {\scriptsize 0.0061} & {\scriptsize 0.0332} & {\scriptsize 0.0374} & {\scriptsize 0.0270}\tabularnewline
\hline 
\end{tabular}

\caption{As table \ref{tab:bad splits}, but for Yule trees. Unexpectedly,
the Jaccard distance performs very well in these simulations. Only
LogDet performs poorly. \label{tab:bad splits Yule}}
\end{table}

\section*{\newpage{}Figures}


\end{document}